\documentclass[sigconf]{acmart}
\pdfoutput=1
\usepackage{booktabs} % For formal tables
\usepackage{float}
\usepackage{supertabular,booktabs}
\usepackage{makecell}
\usepackage[british]{babel}
\usepackage{hhline}
\usepackage{multirow}
\usepackage{tikz}
\usepackage{pgfplots}
\usetikzlibrary{arrows}
\usepackage{euscript}
\usepackage{algorithmicx}
\usepackage[ruled]{algorithm}
\usepackage{algpseudocode}
\usetikzlibrary{arrows, shapes, fit}
\usepackage{tkz-graph}
\usepackage{bbm}
\usepackage{bm}
\usepackage{amssymb}

\makeatletter
\newcommand{\algmargin}{\the\ALG@thistlm}
\makeatother
\algnewcommand{\parState}[1]{\State%
	\parbox[t]{\dimexpr\linewidth-\algmargin}{\strut #1\strut}}

\usetikzlibrary{positioning}
\tikzset{main node/.style={circle,fill=blue!20,draw,minimum size=0.5cm,inner sep=0pt},}
\usepackage{amsmath}

\usetikzlibrary{arrows.meta}
\tikzset{%
	>={Latex[width=2mm,length=2mm]},
	base/.style = {rectangle, rounded corners, draw=black,
		minimum width=4cm, minimum height=1cm,
		text centered, font=\sffamily, text width=4cm },
	activityStarts/.style = {base, fill=blue!30},
	entity/.style = {base},
	workflow/.style={rectangle, draw=black,
		minimum width=2cm, minimum height=0.4cm,
		text centered, font=\sffamily, text width=4cm},
	data/.style={rectangle, rounded corners, draw=black,
		minimum width=2cm, minimum height=0.4cm,
		text centered, font=\sffamily, text width=4cm},
	startstop/.style = {base, fill=red!30},
	activityRuns/.style = {base, fill=green!30},
	process/.style = {base, minimum width=2.3cm, fill=orange!15,
		font=\ttfamily},
}
\DeclareMathOperator*{\argmax}{arg\,max}

\setcopyright{acmcopyright}
\acmDOI{}

\acmISBN{}

\renewcommand\footnotetextcopyrightpermission[1]{} % removes footnote with conference information in first column
\begin{document}
\title{Scalable graph-based individual named entity identification}
%\titlenote{Produces the permission block, and
%  copyright information}
%\subtitle{Extended Abstract}
%\subtitlenote{The full version of the author's guide is available as
%  \texttt{acmart.pdf} document}

\author{Sammy Khalife}
%\authornote{}
%\orcid{1234-5678-9012}
\affiliation{%
  \institution{Ecole Polytechnique, LIX, Dascim}
  \streetaddress{}
  \city{Palaiseau}
  \state{}
  \postcode{F-91120}
}
\email{khalife@lix.polytechnique.fr}

\author{Michalis Vazrigiannis}
%\authornote{}
\affiliation{%
	\institution{Ecole Polytechnique, LIX, Dascim}
	\streetaddress{}
	\city{Palaiseau}
	\state{}
	\postcode{F-91120}
}
\email{mvazirg@lix.polytechnique.fr}

% The default list of authors is too long for headers.
%\renewcommand{\shortauthors}{S et al.}

\begin{abstract}
Named entity discovery (NED) is an important information retrieval problem that can be decomposed into two sub-problems. The first sub-problem, named entity recognition (NER), aims to tag pre-defined sets of words in a vocabulary (called "named entities": names, places, locations, ...) when they appear in natural language. The second subproblem, named entity linking/identification (NEL), considers these entity mentions as queries to be identified in a pre-existing database. In this paper, we consider the NEL problem, and assume a set of queries (or mentions) that have to be identified within a knowledge base. This knowledge base is represented by a text database paired with a semantic graph. We present state-of-the-art methods in NEL, and propose a 2-step method for individual identification of named entities. Our approach is well-motivated by the limitations brought by recent deep learning approaches that lack interpratability, and require lots of parameter tuning along with large volume of annotated data.

First of all, we propose a filtering algorithm designed with information retrieval and text mining techniques, aiming to maximize precision at K (typically for $5 \leq \bm{K} \leq 20$). Then, we introduce two graph-based methods for named entity identification to maximize precision at 1 by re-ranking the remaining top entity candidates. The first identification method is using parametrized graph mining, and the second similarity with graph kernels. Our approach capitalizes on a fine-grained classification of entities from annotated web data. We present our algorithms in details, and show experimentally on standard datasets (NIST TAC-KBP, CONLL/AIDA) their performance in terms of precision are better than any graph-based method reported, and competitive with state-of-the-art systems. Finally, we conclude on the advantages of our graph-based approach compared to recent deep learning methods.
\end{abstract}

%
% The code below should be generated by the tool at
% http://dl.acm.org/ccs.cfm
% Please copy and paste the code instead of the example below.
%
%\begin{CCSXML}
%<ccs2012>
% <concept>
%  <concept_id>10010520.10010553.10010562</concept_id>
%  <concept_desc>Computer systems organization~Embedded systems</concept_desc>
%  <concept_significance>500</concept_significance>
% </concept>
% <concept>
%  <concept_id>10010520.10010575.10010755</concept_id>
%  <concept_desc>Computer systems organization~Redundancy</concept_desc>
%  <concept_significance>300</concept_significance>
% </concept>
% <concept>
%  <concept_id>10010520.10010553.10010554</concept_id>
%  <concept_desc>Computer systems organization~Robotics</concept_desc>
%  <concept_significance>100</concept_significance>
% </concept>
% <concept>
%  <concept_id>10003033.10003083.10003095</concept_id>
%  <concept_desc>Networks~Network reliability</concept_desc>
%  <concept_significance>100</concept_significance>
% </concept>
%</ccs2012>
%\end{CCSXML}
%
%\ccsdesc[500]{Computer systems organization~Embedded systems}
%\ccsdesc[300]{Computer systems organization~Redundancy}
%\ccsdesc{Computer systems organization~Robotics}
%\ccsdesc[100]{Networks~Network reliability}

\keywords{Information retrieval, named entities, graph mining, graph kernels, supervised learning}

\maketitle

\section{Introduction} \label{SEC:INTRO}

%The {\it IJCAI--18 Proceedings} will be printed from electronic
%manuscripts submitted by the authors. These must be PDF ({\em Portable
%Document Format}) files formatted for 8-1/2$''$ $\times$ 11$''$ paper.

\subsection{
	%Definition and illustration
	Basic Concepts }
%The task of \textit{Named entity discovery}
The purpose of \textit{Named entity discovery} (NED) in machine learning and natural language processing is two-fold. First, it aims to extract pre-defined sets of words from text documents. These words are representations of \textit{named entities} (such as names, places, locations, ...). Then, these \textit{entity mentions} paired with their context are seen as \textit{queries} to be identified in a pre-existing database \cite{hachey2013evaluating}.
Firstly, it is important to stress that the subtask of NED - Named entity recognition (NER) - is not trivial since we do not have an exhaustive list of the possible spelling of named entities, moreover their text representation can change (for example, "J. Kennedy" vs. "John Kennedy").

In this paper we focus on the second task, \textit{Named entity linking (NEL)}. Let us define it properly.

\textit{\textbf{Named entity (and Mention/Query)}:} An entity is a real-world object. It usually has a physical existence, but can be abstract. It is denoted with a proper name. In the expression "Named Entity", the word "Named" aims to restrict the possible set of entities to only those for which one or many rigid designators stands for the referent \cite{nadeau2007survey}. When a named entity appears in a document, the words that represent it can also be refered as a \textit{mention}. Finally, a \textit{query} refers to the mention, the context where it appears, and associated type. (We give more explaination on the notion of type in the Knowlege base definition.)

\textit{Example:} "John Kennedy served at the height of the Cold War". In this sentence, John Kennedy is a named entity (or mention), and the associated query is the name "John Kennedy", the sentence, and the named entity type (e.g Person). %. John Kennedy is a senator of the USA, but John Kennedy was also the 35th president of the USA. %How can we distinguish between them using document information ?

\begin{figure}[!h]
	\centering
	
	\begin{tikzpicture}[node distance=1.6cm,every node/.style={fill=white, font=\sffamily}, align=center]
	
	%\tikzstyle{every node}=[font=\tiny]
	\node (entity1)             [entity]              {\baselineskip=10pt \tiny{E1 - PER - John F. Kennedy} ~\\
		\tiny John F. Kennedy is served as the 35th President of the U.S.A \par};
	\node (entity2)     [entity, below of=entity1]          {\baselineskip=10pt \tiny{E2 - ORG - Democratic Party (United States)} ~\\
		\tiny{The Democratic Party is a major contemporary political party in the U.S.A} \par};
	\node (entity3)      [entity, below of=entity2] {\baselineskip=10pt \tiny{E3 - GPE - Washington} ~\\
		\tiny{Washington is the capital of the U.S.A}  \par};
	\draw[->]             (entity1) -- (entity2);
	%\draw[->]             (entity2) -- (entity1);
	\draw[->]			  (entity2) -- (entity3);
	\draw[->]			  (entity3) -- (entity2);
	%\node[xshift=1.2cm,yshift=-1.5cm, text width=2.5cm]{The activity comes to the foreground}(onResumeBlock.east);
	\end{tikzpicture} 
	
	\caption{\label{knowledgebase} Representation of a \textit{unweighted directed semantic graph} (Wikipedia/NIST TAC-KBP Challenge 2010). An edge between two entities $E_1$ and $E_2$ represents a link from $E_1$ to $E_2$.}
\end{figure}
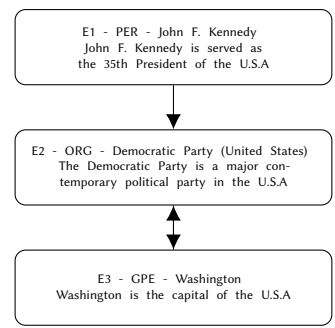

\textit{\textbf{Knowledge base/graph :}} A Knowledge base is a database providing supplementary descriptive and semantic information about entities. The semantic information is contained in a knowledge graph, where a node represents an entity, and an edge represents a semantic relation. In the general case, the knowledge graph can be of any kind (directed, weighted, ...). See figure \ref{knowledgebase} for an example. We discuss knowledge graph types in details in part 3 and Evaluation.

Usually, entities have a type feature \cite{nadeau2007survey}, such as : \textit{PER, ORG }and \textit{GPE}  (respectively \textit{person, organization} and \textit{localization}). For instance, the entity "United States" is a GPE, John Kennedy is a PER, etc... These types play a central role in NER and NEL. Indeed, these features decrease the number of candidate entities for identification.
%(** why? what kind of role?). 
However, it is possible to consider a fine-grained classification, with hundreds of entity types, similarly to DBPedia ontology\footnote{http://wiki.dbpedia.org/services-resources/ontology}. 

\textit{\textbf{Named entity linking  (NEL)}:} Given mentions of entities in digital data (\textit{named entities}, or \textit{mentions}) the purpose of named entity linking is to identify the corresponding \textit{unique} (one entity per mention) ground truth entities (also referred as \textit{gold entities}) in a database (\textit{knowledge base}).

\subsection{Contributions}
In this work, we provide  an overview of the \textit{NEL} problem, and investigate two graph-based methods. We highlight their advantages and limitations over recent deep learning approaches. %Moreover, our approach is well-motivated for applications of NEL (social networks, or compliance processes), 

In the following, the first step, refered as \textit{entity filtering}, aims at reducing entity candidates to top $\bm{K}$ entities for one \textit{query}. The second step, refered as \textit{entity identification}, aims at identifying the true entity among the remaining $\bm{K}$ candidates, for which we propose two graph-based routines (selective graph mining and graph comparison with kernels). We present the construction of our algorithms in details, along with their computational complexity. We also include an evaluation of experimental performance on several datasets, with an analysis of the impact of parameter $\bm{K} \in \{5, ..., 20\}$, and a detailed comparison with existing approaches. We do not include in this work the problem of \textit{Fine-grained named entity recognition} \cite{ling2012fine} (word tagging) nor \textit{NIL-detection} (detection of entities out of the knowledge base).

\section{Related work} \label{sec:relatedWork}
%named entity linking 
%A NEL algorithm implies having a \textit{Knowledge Base (KB)}, which is considered constant throughout our algori . 
In general, linking between named entities and a knowledge graph can be done either \textit{individually} or \textit{collectively}. In the first case individual mentions are considered separately, as independent queries. In the collective way, we consider dependence between queries in a same document, where the true entities should have some proximity, or coherence. Therefore, a collective linking framework implies some dependance between these entity variables. For the sake of completeness, we report here both individual and collective approaches. In the following subsections, we present 3 categories of state-of-the-art algorithms for named entity linking. 

%There are 3 categories of state of art NEL approaches that will be presented in the following subsections.

%\subsection

\textit{\textbf{Notations :}} $\bm{E} = \{1, ..., E\} \subset \mathbb{N} $: indexes of entities  and $ \bm{M} = \{1, ..., M\} \subset \mathbb{N} $: indexes of mentions, $\hat{e}_{i}$: system's output entity index for mention index $m_i$.

\subsection{Graphs for NED} \label{ssec:Graphs}

%\begin{itemize}
%\textcolor{blue}{\textbf{Input : }} Graph with reference entities and mentions, edges built on context similarities ~\\
%  \centering \includegraphics[width = 6.5cm,height = 4.5cm]{dream.png}	
%Node a and b are the mentions
%Node 1,2,3

\textit{\textbf{Individual \& collective linking :}} Given a real value scoring function defined on the product space of mentions and entity states (for example, combinations of Jaccard index over N-grams), let $W_{i, j}$ the corresponding score between the mention $i$ and the entity $j$. For individual disambiguation, one wants to perform independent mention-entity attribution. Then the graph structure is irrelevant and the formulation is straightforward :
\begin{equation} \label{eq:individualCase}
\hat{e_i} = \argmax_{j \in E} W_{i,j}
\end{equation}

In the collective linking formulation, the optimization formulation is different : the underlying \textit{gold entities} should respect some arbitrary semantic coherence. The coherence information is represented within a \textit{coherence function} $\psi : \bm{E}^{M} \rightarrow \mathbb{R}$ between the entity candidates. Usually $\psi$ is defined using knowledge graph structure. For example $\psi$ can be defined as the shortest-path function on the \textit{knowledge graph}. With these notations, the set of selected entities are formally defined as :
\begin{equation} \label{eq:collectiveCase}
\hat{e_1}, ..., \hat{e_{M}} = \argmax_{j_1, .. , j_m \in E^M} [  ( \sum_{l=1}^{M} W_{l, j_{l}} ) +  \psi(j_1, ..., j_M) ]
\end{equation}

Equation \ref{eq:collectiveCase} can be formulated as a boolean integer program, but the nature of $\psi$ being arbitrarly complex (e.g shortest-path function) does not allow to solve the general case, especially when $M \rightarrow +\infty$. Therefore, other formulations are preferred : a rule-based individual linking has been proposed \cite{guo2011graph}, and \cite{han2011collective} proposed a collective formulation for entity linking decisions, in which evidence can be reinforced into high-probability decisions. 

Other formulations using Community detection and Pagerank have been proposed, for which we give detailed explainations in the next paragraphs.

\begin{figure}[!h]
	\centering
	\begin{minipage}[c]{0.4\linewidth}
		\begin{tikzpicture}
		\begin{scope}[xshift=4cm]
		\node[main node] (1) {$a$};
		\node[main node] (2) [below = 0.5cm  of 1] {$b$};
		\node[main node] (3) [right = 1cm  of 1]  {$e_1$};
		\node[main node] (4) [right = 1cm  of 2] {$e_2$};
		\node[main node] (5) [below = 0.5cm  of 4] {$e_3$};
		%\node[ellipse,draw=green, fit=(3) (5)](KB) {};
		\node[ellipse, draw=black,fit=(1) (2)](Mentions) {};
		\node[ellipse, draw=black,fit=(3) (4) (5)](KB) {};
		
		\draw[<->] (1) -- (3);
		\draw[<->] (1) -- (4);
		\draw[<->] (1) -- (5);
		\draw[<->] (2) -- (3);
		\draw[->] (3) -- (4);
		\draw[<->] (4) -- (5);
		\draw[<->] (2) -- (4);
		\draw[<->] (2) -- (5);
		%\path[draw,thick]
		%(1) edge node {} (3)
		%(2) edge node {} (4)
		%(2) edge node {} (5)
		
		%(4) edge node {} (5)
		%(5) edge[bend right] node {} (3);
		
		\end{scope}
		\end{tikzpicture}
	\end{minipage}
	\begin{minipage}[c]{0.4\linewidth}
		\begin{tikzpicture}
		\begin{scope}[xshift=4cm]
		\node[main node] (1) {$a$};
		\node[main node] (2) [below = 0.5cm  of 1] {$b$};
		\node[main node] (3) [right = 1cm  of 1]  {$e_1$};
		\node[main node] (4) [right = 1cm  of 2] {$e_2$};
		\node[main node] (5) [below = 0.5cm  of 4] {$e_3$};
		%\node[ellipse,draw=green, fit=(3) (5)](KB) {};
		\node[ellipse, draw=black,fit=(1) (2)](Mentions) {};
		\node[ellipse, draw=black,fit=(3) (4) (5)](KB) {};

		\path[draw,thick]
		(1) edge node {} (3)
		(1) edge node {} (4)
		(1) edge node {} (5)
		(2) edge node {} (3)
		(2) edge node {} (4)
		(2) edge node {} (5)
		(3) edge node {} (2)
		(3) edge node {} (4)
		(4) edge node {} (5);
		%(4) edge node {} (5)
		%(5) edge[bend right] node {} (3);
		
		\end{scope}
		\end{tikzpicture}
	\end{minipage}
	
	\caption{Directed and undirected query/entity bipartite weighted graph. For each graph, the set on the right is built using an extract of the \textit{knowledge graph} (same extract as figure \ref{knowledgebase}) : Nodes $e_1, e_2, e_3$ are entities the knowledge base. Nodes $a$ and $b$ are entity mentions extracted from text documents forming queries. } \label{fig:2}
\end{figure}
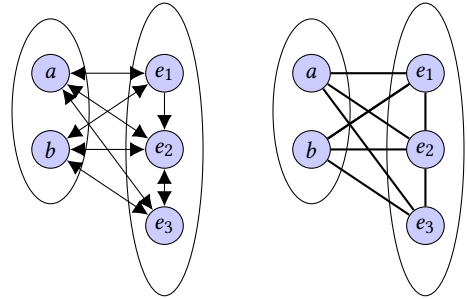

\textit{\textbf{Bipartite graph \& community detection : }} Similarly, we can model \textit{NEL} as a bipartite graph optimization problem. One of the nodes set is built using the knowledge graph : the graph can be directed or undirected, and weighted (using similarity functions for instance). In the entity nodes set, the information is structured, clear, canonic and considered true. The second nodes set are the \textit{queries} and contain potentially ambiguous information (cf figure \ref{fig:2}). 
%s contained for queries representing the same entity.

In this context, graph-based approaches have been developed. \cite{hoffart2011robust}, and \cite{alhelbawy2014graph} proposed to link efficiently mentions to their corresponding entities using the weighted undirected bipartite graph built among mentions-entities text similarities, by extracting a dense subgraph in which every mention node is connected to exactly one entity, yielding the most likely disambiguation.

In general, this combinatorial optimization problem is \textit{NP-hard} with respect to the number of nodes, since they generalize Steiner-tree problem \cite{hoffart2011robust}. However, heuristics have been brought forward, such as \cite{hoffart2011robust} and \cite{alhelbawy2014graph} proposing a discarding algorithm using taboo search and local similarities with polynomial complexity.

\textbf{\textit{ PageRank :}} Another graph-based proposal is to use an adaptation of PageRank algorithm to provide entities a popularity score. For example, \cite{usbeck2014agdistis} built a weighted graph G of all mentions and entities based on local and global similarities, and capitalize on the Hyperlink-Induced Topic Search (HITS) algorithm to produce node authority scores. on the graph. Then, within similar entities to mentions, only entities with high authority will be retained.

These graph-based algorithms proved to be fast compared to the other approaches since they do not require any training, and perform reasonably well in terms of precision. However, most of them are greedy and need pre-filtering to discard some entity candidates.
%- PGMs ~\\

\subsection{Probabilistic graphical models} \label{PGM-NEL}
An interesting idea is to consider mentions as random variables and their golden/true entities as hidden states. Unlike character recognition where $|E| = |E_i| = 26$ for latin alphabet, the number of possible states per entity - usually $ \geq 10^6$ - and Viterbi algorithm quadratic complexity ($ \mathcal{O}(N \left|{S}\right|^{2}) $, where $ S $ is the number of states and $N$ the number of observations) makes the problem computationally untractable. To overcome this technical issue, a first step proposed by \cite{alhelbawy2013named} is to establish a reduced set of candidates per mention : $m_i \in E_i$ using mention context. Using annotation, an HMM is trained on the reduced set of candidates. Inference is made using message passing (Viterbi algorithm) to find the most probable named entity sequence. Another approach using probabilistic graphical model has been provided by \cite{ganea2016probabilistic}, with a factor graph that uses popularity-based prior, Bethe Approximation to decrease inference computational cost, and message passing to compute marginal probabilities. The computational complexity is $\mathcal{O}(N^2  r^2)$ where $r$  the number of average entity candidates per mention and $N$ the number of observations.

Finally, another probabilistic graphical model has been proposed, similarly to latent Dirichlet allocation (LDA) \cite{blei2003latent}, where an iterative procedure $P$ is used above the LDA-scheme to enrich the knowledge base. Its complexity is proportional to the product between LDA complexity and the number of iterations of procedure $P$ \cite{Li:2013:MEN:2487575.2487681}.
%	\item[] \textbf {Evidence mining for NED }\cite{Li:2013:MEN:2487575.2487681} 

\subsection{Embeddings and deep architectures}

Word embeddings are practical and used for deep learning architectures \cite{goldberg2014word2vec}. Methods such as Word2vec and Glove build a statistical distribution over words representations scalar products \cite{goldberg2014word2vec,pennington2014glove}. Considering these pairwise conditional probabilities, Skip-gram model aim is to predict context words given one input word $w_{i}$.
Indeed, each word has a probability of appearing given words around it, with a probability being a growing function of the dot product between context word vectors representations. 
These word embeddings can be obtained either outside or inside of a deep learning architecture, as a first layer. Here, the embeddings represent words of mentions context and entities text description. An example of learning other representations entities is achieved in \cite{yamada2017Learning} and reaches state-of-the-art performance on NIST TAC-KBP 2010 Dataset.

A disambiguation tool using pre-trained embeddings, then averaging and ranking has been proposed \cite{yamada2016joint} with a $ O(m  e^{2}) $ complexity, where $m$ is the number of mentions and $e$ the number of entities.

Recent advances in neural networks conception suggested to use word embeddings and convolutional neural networks to solve the named entity linking problem. \cite{sun2015modeling} proposed to maximize a corrupted cosine similarity between a mention, its annotated gold entity and a false entity. The network is trained with polynomial complexity, and reached state-of-art performance in precision (until 2017 and \cite{yamada2017Learning}) on NIST TAC-KBP datasets in 2009 and 2010.

Long-short-term memory networks (LSTMs) recently provided remarkable results for natural language modeling in general. Recent neural network architecture have been proposed \cite{sil2017neural,raiman2018deeptype}, the latest using a recent method using fine-grained ontology type system and reaching promising results on several datasets.
%\subsection{Format of Electronic Manuscript}

%\begin{quote} 
%\mbox{\tt $\backslash$usepackage\{times\}}
%\end{quote}

%in the preamble.\footnote{You may want also to use the package {\tt
%latexsym}, which defines all symbols known from the old \LaTeX{}
%version.}
%\subsection{C}

\subsection{Comments}
\begin{figure}[!h]
	\centering
	\begin{tikzpicture}[node distance=1cm,every node/.style={fill=white, font=\sffamily}, align=center]
	%\tikzstyle{every node}=[font=\tiny]
	\node[data] (work1)                          {\baselineskip=10pt \tiny{Input Knowledge base and extracted mentions}  \par};
	\node (work2)		[workflow, below of=work1]	{\baselineskip=10pt \tiny{Entity filtering} \par};
	\node (work3)		[workflow, below of=work2]	{\baselineskip=10pt \tiny{Entity linking\\ (Ranking, regression or classification)} \par};
	\node (work4)		[data, below of=work3]	{\baselineskip=10pt \tiny{Output} \par};
	\draw[->]             (work1) -- (work2);
	\draw[->]			  (work2) -- (work3);
	\draw[->]			  (work3) -- (work4);
	%\node[xshift=1.2cm,yshift=-1.5cm, text width=2.5cm]{The activity comes to the foreground}(onResumeBlock.east);
	\end{tikzpicture}
	\caption{Visualization of a NEL workflow. Entity linking is performed in 2 steps.}
	\label{NELWorkflow} 
\end{figure}
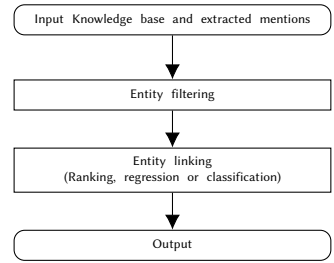

Each of the cited approaches uses a filtering metric to discard non relevant entities. To the best of our knowledge, this is the case for every state-of-the-art routine. Therefore, the final precision score will be upper-bounded by the recall of the filtering, shown on figure \ref{NELWorkflow}. 
it is widely accepted that neural networks require very large datasets. Moreover, it is data quality of automatic generated mentions from Wikipedia is debatable. 

%In our work, only individual linking is considered : our linking system is applied on each query separately. The next subsection is dedicated to the entity filtering procedure. 

\section{Methodology}
Due to the aforementioned comments, we chose to investigate graph-based methods along with supervised learning algorithms requiring a reasonable amount of data. 
This section addresses entity filtering and identification. In the first subsection, we explain the experimental factors causing a drop in performance for entity filtering. Based on this analysis, we propose a new entity filtering method using information retrieval techniques. In the second subsection, we present two new graph-based methods for entity identification.

All our method is conceived for individual linking : queries are considered separately.
\subsection{Entity filtering}\label{sec:section3Entity}

Assume the following document sentence: "J. Kennedy has met soviet Premier Nikita Khrushchev and Kennedy in 1961, in Vienna". A named entity recognizer will tag "J. Kennedy" as a mention. We would like to discard wrong entity candidates using named entity filtering. To do so, we consider only two sources of information in the query : the mention name ("J. Kennedy"), and the information contained in the rest of document, i.e the other words. The filtering precision is an upper bound of the overall precision of the NEL workflow displayed in figure \ref{NELWorkflow}.
% In this section we report results of experiments based on queries using only name information. %We stress out the knowledge base used for our entity filtering experiments is a XML dump of Wikipedia provided by NIST TAC-KBP challenge (refer to section Experiments for more details).

\subsubsection{Example of entity filtering}~\\
Given a filtering algorithm $\bm{F}$, let $\bm{E}_{\bm{G}_M}$ be the generic golden entity random variable of mention $\bm{M}$. Let $\bm{p_F}$ be the following mention/entity prior :

\begin{equation} \label{eq:priorFiltering}
\bm{p_F} =\mathbb{P}(\bm{E}_{G_{\bm{M}}} \in \bm{F}(\bm{M}))
\end{equation}
A performant filtering method should maximize this prior. In information retrieval, and especially for NEL, one of the most popular metric is built using combinations of N-grams. \cite{han2011collective,ferragina2010tagme,blanco2015fast,nguyen2016j}. Considering these sets of N-grams, the Jaccard index is a real value defined by the ratio between the size of the intersection of these sets divided by the size of the union of the sets. Let us denote $\bm{F_{\mathcal{N_{K}}}}$ the filtering algorithm (here $\bm{K}$ top results) using Jaccard index score using n-grams. We suppose the type of the query is known in advance (PER, ORG, GPE), and consider entity candidates accordingly.
%, and W capturing language specificites, i.e acronyms, and nicknames.
%In the next subsection, we propose a method in order to capture $W_M$  .

We compute the prior parameter estimation (here $n$ is the number of samples):

\begin{equation} \label{eq:empiricalPriorFiltering}
\bm{\hat{p}_{F_{\mathcal{N_{K}}}}} = \frac{1}{n} \sum_{i=1}^{n} \mathbbm{1}_{\{\bm{E}_{G_{\bm{M}_{i}}} \in \bm{F}_{\bm{\mathcal{N_{K}}}}(\bm{M_i})\}}
\end{equation}

over 2010 NIST TAC-KBP dataset (we refer to experiments section for a description of the datasets).
Using mean of Jaccard index over 2-grams, 3-grams and 4-grams, and $\bm{K} = 50$ , it turns out that $\bm{\hat{p}_{F_{\mathcal{N_{K}}}}} = 0.83$. This implies that one cannot hope to reach a better empiral precision at 1 that $83\%$ if we chose this filtering method. A low value of $\bm{p_{F_{\mathcal{N_{K}}}}}$ causes a greedy drop of precision. Therefore, it is important to maximize $\bm{R}@\bm{K}$ in the first place. 
%to use $\bm{F_{\mathcal{N}}}$ filtering.
% Github repository\textsuperscript{\ref{githubNelNote}}

\subsubsection{Experimental explanation}~\\
In this subsection we emphasize factors causing drop in precision with the previous filtering example.

\textbf{\textit{Acronyms :}}
%By looking on mentions mis-identified, 
Acronyms played an important role in entity filtering. For example, in NIST TAC-KBP 2009 dateset, "gsu" represents Georgia state university, "ccp" is used for Communist party of China, "abc" for Australian broadcasting corporation, "cdu" for Christian Democratic Union (Germany).

\textbf{\textit{Nicknames, other names, ancient names :}}
Geographical entities such as cities and countries have some historical background, and can be referenced by a query that is not similar to the ground truth entity name. For example, "Beehive state" refers to Utah state in the United State of America, "Flavia Neapolis" or "Little Damascus" or "Shechem" refer to Naplouse in Cisjordanie, "Garden city" to Port Harcourt in  Nigeria. Other entity types such as people are concerned : "Iron Lady" refers to Margaret Thatcher, ex-Prime Minister of United Kingdom.

\textbf{\textit{String comparison is not enough : }}
String comparison sometimes fails due to overlapping of mentions and entities names. This becomes a problem when the number of entities in the knowledge base is high. For example, Wikipedia knowledge base contains millions of entities (cf. experiments section). In this case, Jaccard distance over combinations N-grams cannot capture resolve ambiguity. For example, "State of Utah" has as 1st ranked entity: "Nevada State Route 531", whereas its gold entity is "Utah".
%Homonymy and one of the three situations (acronyms, nicknames and overlapping surface forms) can happen simultaneously. For instance different surface form and homonymy with "Iron Lady" for Margaret Thatcher and Biljana Plavšić, ex-president of Serbia).

\newcommand\given[1][]{\:#1\vert\:}

\subsubsection{Our filtering method :}~\\
In order to improve previous filtering algorithms, we propose a routine based on four main components : preprocessing, acronym detection and expansion, name scoring and context scoring. In practice, pre-processing is applied first on the data. Then, the three remaining steps (gathered in algorithm \ref{Algorithm:scoring}) are performed.

\textbf{\textit{a - Preprocessing : }} 

For trivial queries having a mention name equal to an existing entity name, we implemented a naive match pre-processing. This is performed by saving a dictionnary which keys are the names of knowledge base entities, and values the actual entity ID. 

\textbf{\textit{b - Acronym detection/expansion}}~\\
(refered as $\bm{ACRD}$ in algorithm \ref{Algorithm:scoring}).

Acronym detection and expansion is a common topic in bioinformatics. We refer to \cite{ehrmann2013acronym} as a survey of acronym detection methods.
\begin{itemize}
	\item Detection : Following \cite{zhang2011entity}, we explored a supervised learning approach, but chose simple rule-based decision based on the string length and cumulated length between each capital letter \cite{gusfield1997algorithms}.
	\item Expansion : The score is the length of longest common substring \cite{Apostolico1987} between acronym string and capital letters of the entity target
\end{itemize} 

\textbf{\textit{c - Name scoring}} (Refered as JN in algorithm \ref{Algorithm:scoring})

When the named entity (mention) is not tagged as an acronym, comparison with entity titles is performed :
\begin{itemize}
	\item Compute  $\mathcal{N}-$grams  for $N \in \{2, 3, 4\}$ 
	\item Average Jaccard Index of mention name and entity title. [definition of Jaccard index in 3.1.1]
\end{itemize}

\textbf{\textit{d - Context scoring : }} (refered as \textit{tfidfScore} in algorithm \ref{Algorithm:scoring})

This part aims to maximize similarity of sentences. We experimented several techniques :
\begin{itemize}
	\item Keyword detection (by frequency) and intersection
	%\item Average of cosine similarity of word embeddings
	\item Document distances using word embeddings, especially Word Mover's Distance \cite{kusner2015word}
	\item Cosine similarity of TF-IDF vectors. The matrix is computed on all the knowledge base. It is refered as \textit{tfidfScore}, and  The scoring algorithm is summarized in Algorithm \ref{Algorithm:scoring}. 
\end{itemize}

We decided to keep cosine similarity of TF-IDF vectors for three reasons. First, for experimental reason : we did not see any significative difference in performance using word embeddings. Second, this choice seemed more consistent with the motivation of this paper to challenge recent approaches using such embeddings).  
Finally, we wanted to propose a straightforward candidate generation method that can be implemented in a real system, without any annotated training data.

The scoring algorithm for entity filtering is summarized in Algorithm \ref{Algorithm:scoring}. $\hat{T}$ is the named entity considered, given as input in the data.  Its linear computational complexity with respect to the number of queries and entities is immediate (reminded in part 3 and 4). (All the code is included in our code repository (available on demand)). 

%on Github\footnote{\label{githubNelNote}https://github.com/Khalife/ker-nel-experiments}.
\alglanguage{pseudocode}
\begin{algorithm}[!h]
	\small
	\caption{Entity filtering (generation of entity candidates)}
	\label{Algorithm:scoring}
	\begin{algorithmic}[1]
		\Require{Parameter $\bm{K}$, Query $ ( \bm{Q} = (\bm{M}, \bm{C}, \bm{\hat{T}}))$, Entities $(\bm{E}_j, \bm{T}_j)_{1 \leq j \leq \bm{E}}$,}
		
		%\State $ \bm{DS} \leftarrow \{\}$ 
		%\For {$i = 1 \to M$}
		%\State Perform preprocessing
		\State $\bm{ds} = [\,]$
		
		\State $ \bm{y}_{acr} \leftarrow \bm{ACRD}(\bm{M}_{i})$ 
		\For {$ j = 1 \to E$  }
		\If{$T_j  == \hat{T}$}
		\If{ $\bm{y}_{acr} == 1$}
		\State $s_n = acronymScore(\bm{M}_{i}, \bm{B}_j)$    
		\Else
		\State $s_n =  JN(\bm{M}_{i}, \bm{B}_j) $ 
		\EndIf
		%\State $ s_c = tfidfScore(\bm{C}_{i}, \bm{B}_j)$
		\State $ s_t = \frac{1}{2}(s_n + tfidfScore(\bm{C}_{i}, \bm{B}_j)) $
		\State Sorted insertion by value of $\{j : s_t\}$ in $\bm{ds}$ 
		\EndIf
		
		\EndFor	
		%\State $\bm{DS}[\bm{M}_i]  \leftarrow  \bm{ds}$ 
		%\EndFor	
		%\State \Return $\bm{DS}$ (Mention to entity scores)
		\State \Return $\bm{ds[:\bm{K}]}$ ($\bm{K}$ top entities )
		\Statex
		
	\end{algorithmic}
	%\vspace{-0.4cm}%
\end{algorithm}

%\vspace{0.5cm}%

\newpage
\subsection{Graph-based ranking algorithms}\label{sec:framework}
In this section, we propose two different graph-based procedures for \textit{named entity identifcation}. We capitalize on our filtering method (cf previous subsection) to maximize precision at $\bm{K}$ (we denote it $\bm{R}@\bm{K}$, typically with $ 5 \leq \bm{K} \leq 20$), in order to get a limited amount of entity candidates. These graph-based methods use enriched features extracted from the knowledge graph, and re-rank these top entity candidates in order to return the ground truth as first ranked entity. %The role of entity types is central in both of our methods. %More details on how this fine-grained type features have been constructed are presented in the experiments section.
% All of our source code with documentation is available on Github\textsuperscript{\ref{githubNelNote}}.

\subsubsection{Knowledge graphs structures}~\\

Our methods do not take in account weights for two reasons. 

First, by definition : edges indicate a semantic relation, and it is difficult to estimate the intensity of such relation in practice. One could think of this intensity as a quantity proportional to the number of occurences of two entities together, but we do not have access to such information in the knowledge base, where links can appear only once but still be significant (Example : Country to its City capital). A natural way to assign an intensity to each relation would use edge classification, but we did not include it in this paper.

Second, our methods precisely allow graph search independently of edges weights. Similarly, our algorithms do not penalize local search due to global properties of the graph, on the contrary of methods such as PageRank that assign each node a popularity.

Through the experiments of the paper, we assumed unweighted and undirected knowledge graph eventhough our methods can also be applied to directed graphs.

\subsubsection{First method :  parametrized graph mining }~\\
A natural idea to take into semantic information is to use graph mining on the knowledge graph. To do so, we propose for one entity candidate to aggregate context scores of relatively \textit{"close"} nodes in the knowledge graph. This context scores are computed similarly as for the filtering method (TF-IDF cosine similarity).
%The definition of \textit{"close"} in the graph can be direct neighbors, or nodes at distance lower or equal than a threshold. 

\textbf{\textit{Graph \& node neighborhood :}} There are several ways to define a node neighborhood in a graph. We implemented two situtations : first, direct neighbors, and then using breadth-first-search (BFS) until a distance threshold. We did not see any improvement with the second method, we present here the method with direct neighbors. %(cf comments subsection).
%In-depth (**depth in the graph? you were talking on BFS and now you say dept. ...please explain) 
Graph exploration creates a sparsity-noise trade-off (sparsity in case of a low numbers of new entities, and noise from irrelevant entities).
%Entity candidate degree 
%The candidate entity degree has obviously an important impact on the new score features.

%For a given entity candidate $e_j$, if $deg(e_j) < |T|$, then we complete scores by $0$ or use graph exploration to complete the score vector.

To control this trade-off and select a \textit{"convinient"} set of entity neighbors, we parametrize this neighborhood selection using a type-mapping function. More formally, let $T$ be the set of indexes of ontology types. A type-mapping function is a (symmetric) boolean function defined on couples of entity types :
%$ T \rightarrow T \times T \times ... \times T$
\begin{alignat*}{2}
\varphi \in \Phi:   T \times T & \longrightarrow \{0, 1\} \\
%(\mathbf{t_1}, \mathbf{t_2})&\longmapsto \phi(\mathbf{t_1}, \mathbf{t_2})
(t_1, t_2)&\longmapsto \phi(t_1, t_2)
\end{alignat*}
Therefore, the property $\phi(t_1, t_2) = \phi(t_2, t_1)$ should be always true. By definition, $\phi(t_1, t_2) = 1$ if and only if type 1 and type 2 are "jointly" interacting.  i.e given one type or the other, the second has to be considered through graph mining.

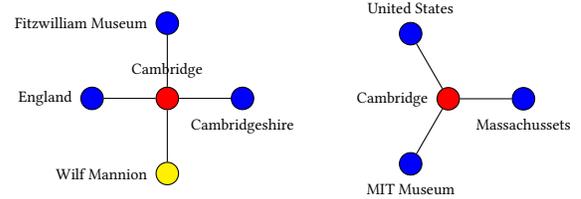
\begin{figure}[!h]
	
	\begin{minipage}[c]{0.4\linewidth}
		\begin{tikzpicture}[every node/.style={sibling distance=30mm}]
		\def \radius {1cm}
		\node[draw, circle, fill=red, label=above:\scriptsize Cambridge] at (360:0mm) (center) {};
		
		\node[draw, fill=blue, circle,label=below:\scriptsize Cambridgeshire] at ({0*90}:\radius) (0) {};
		\draw (center)--(0);
		\node[draw, fill=blue, circle,label=left:\scriptsize Fitzwilliam Museum ] at ({1*90}:\radius) (1) {};
		\draw (center)--(1);
		\node[draw, fill=blue, circle,label=left:\scriptsize England] at ({2*90}:\radius) (2) {};
		\draw (center)--(2);
		\node[draw, fill=yellow, circle,label=left:\scriptsize Wilf Mannion] at ({3*90}:\radius) (3) {};
		\draw (center)--(3);
		
		\end{tikzpicture}
	\end{minipage}
	\hspace{1cm}
	\begin{minipage}[c]{0.4\linewidth}
		\begin{tikzpicture}
		\def \radius {1cm}
		\node[draw, circle, fill=red, label=left:\scriptsize Cambridge] at (360:0mm) (center) {};
		%\foreach \i  in {0,...,2}{
		\node[draw, fill=blue, circle,label=below:\scriptsize Massachussets] at ({0*120}:\radius) (0) {};	
		\draw (center)--(0);
		\node[draw, fill=blue, circle,label=above: \scriptsize United States] at ({1*120}:\radius) (1) {};	
		\draw (center)--(1);
		\node[draw, fill=blue, circle,label=below:\scriptsize MIT Museum] at ({2*120}:\radius) (2) {};	
		\draw (center)--(2);
		
		%}
		\end{tikzpicture}
	\end{minipage}
	\caption{\label{Cambridge} Two homonyms : Cambridge cities. \\
		Left : England city. Right : Massachussets, United States. Following equation \ref{eq:cambridgeEquation}, a neighbor entity $e$ is in blue if $\phi(T_{\text{City}}, T_\text{e}) = 1 $ and yellow otherwise (Wilf Mannion is a former Cambridge football Player in England).}
\end{figure}

\textbf{\textit{Example : Cities :}}
In this example indexes from 0 to 4 represent respectively entity types "City", "State", "Museum", "Country", and "FootballPlayer. Let us choose a mapping function such that:
\begin{equation} \label{eq:cambridgeEquation}
\phi(0, t_1)= 
\begin{cases}
0 & \text{if } t_1 = 4 \\
1          & \text{otherwise}
\end{cases}
\end{equation}
i.e we do not consider type interaction between City and FootballPlayer in the knowledge graph. We suppose that information based on country, state museum helps to disambiguate cities [cf figure \ref{Cambridge}].

%\begin{figure}[!h]
%	%\centering
%	\hspace{-1.5cm}
%	\begin{minipage}[c]{0.4\linewidth}
%		%	\hspace{-0.5cm}
%		%\centering
%		\includegraphics[height=4cm,width=1.3\linewidth]{cambridge.png}
%		%\caption{Picture 1}
%		\label{fig:1AC}
%	\end{minipage}
%	\hspace{0.3cm}
%	\begin{minipage}[c]{0.4\linewidth}
%		\centering
%		\includegraphics[height=4cm, width=1.3\linewidth]{cambridge-US.png}
%		%\caption{Picture 2}
%		\label{fig:2AC}
%	\end{minipage}
%	\caption{\label{Cambridge} Two homonyms : Cambridge cities. To the left, in England. To the right, in Massachussets, in the United States. With $T_{city}$ the index of City type, a neighbor entity $e$ is in blue if $\phi(T_e, T_{city}) = 1 $ and yellow otherwise (Wilf Mannion is a former Cambridge football Player in England).}
%\end{figure}

\textbf{\textit{Features extraction :}} 
%defined on all ontology types, providing an optimal potentially supplementary entities.
Let $q$ and $e$ respectively a query and one entity. For the sake of simplicity, we denote by $X^{q, e}$ the generic features vector associated with the couple $(q,e)$. $(n_i)_{1 \leq i \leq deg(e)}$ represents the neighbors of e, $t_q$ and $t_{n_i}$ respective entity types,  and a scoring function $s$ (for example, \textit{tfidfScore} in algorithm \ref{Algorithm:scoring}). With same notations, we define the corresponding features as follows :
\begin{equation} \label{eq:graphFeatures}
\forall i \in \{1, ..., |T|\}, ~ (X^{q, e})_i= 
\begin{cases}
s(q, e)& \text{if } i = 1\\
s(q, n_{i}) \varphi(t_q, t_{n_i})             & \text{otherwise}
\end{cases}
\end{equation}
For a given entity candidate $e_j$, if $deg(e_j) < |T|$, then we complete scores by $0$ or use graph exploration to complete the score vector. Here, function $\varphi$ plays the role of a hyperparameter. 

A sample class is defined as : \begin{equation}\label{eq:graphLabels}
Y^{q,e}= 
\begin{cases}
1 & \text{if e is the gold entity}\\
0  & \text{otherwise}
\end{cases}
\end{equation}
\textbf{\textit{Supervised NEL training :}}  With this formulation, we can train \textit{NEL} regressors or classifiers in a supervised learning framework. At inference, the couple $(q, \hat{e})$ maximizing the prediction score yields predicted entity $ \hat{e}$. If same scores are returned for different couples, we return the first candidate. (This situation didn't occur in practice). For our experiments, we used simple classifiers : regression trees, random forests, and logistic regression (details in experiments section).

\textbf{\textit{Hyper-parameter tuning :}} Using a boolean formulation to represent hyper-parameter function $\varphi$, %(\ref{eq:mapOptimization}) 
its selection can be interpreted as a boolean combinatorial optimization problem. The empirical optimal mapping depends a-priori on the knowledge graph structure \textit{and} the queries. The evaluation of the training cost function is not "immediate" since one has to extract new-features for each new mapping function. 

This challenge combined with a tough combinatorial problem invited us to consider the following "routine" :
\begin{itemize}
	\item Train \textit{NEL} systems with trivial $\varphi$ (constant equal to 1)
	%\item Display  mis-ranking by entity types
	%\item Extract new graph features only for a set $\tilde{T}$ of most frequent ontology types concerned by identification-errors [for instance Cities, for more details, cf experiments section]. This is equivalent to constraint $\phi$ to a subset of mapping functions $\{\phi \in \Phi \: | \: \phi(t_{i}, t_{j}) = (0, .., 0) \quad \forall i, j \in \tilde{T} \times \tilde{T}  \}$ 
	\item Use meta-heuristics methods for hyperparameter optimization, mainly genetic programming \cite{Goldberg:1989:GAS:534133}.
\end{itemize}

\textbf{\textit{Graph mining and inference procedure :}} Based on the previous statements and Figure \ref{Cambridge}, we sum up graph mining routine in algorithm \ref{Algorithm:graph-mining-scoring}. The final inference routine is presented in algorithm \ref{Algorithm:KER for NEL}.

\begin{algorithm}[!h]
	\small
	\caption{Selective graph mining (SGM) method }
	\label{Algorithm:graph-mining-scoring}
	\begin{algorithmic}[1]
		
		\Require{Knowledge Graph $\bm{G}$, Query $\bm{Q}$, Entity top candidates with initial filtering score $( \bm{E}_{i}, \bm{S}^{i}_{0} )_{1 \leq i \leq K}$,  Projection types $(\bm{T})_{1 \leq j \leq |T|}$, Disambiguation map $\bm{\phi}$}
		%\State $\bm{C}_{0} \leftarrow \emptyset$
		%\State $\hat{\bm{H}}_{0} \leftarrow \emptyset $
		\For{$ i = 1$ to $K$}
		\parState{Get neighbor nodes in the graph $(\bm{\EuScript{N}}_{j}(\bm{E_{i}}))_{1 \leq j \leq |T|}$ in G according the the type mapping $\phi$}
		\parState{Build scores vector $ \{\bm{S}^{i}_{0}\} 	\cup	\{{\bm{S}^{i}_{\bm{\EuScript{N}}_{j}}}_{1 \leq j \leq |T|}\}$ for each type according to equation \ref{eq:graphFeatures}}
		%node using scoring TF-IDF comparison between entity description and the query $\bm{Q}$}
		\EndFor
		%\State Rank nodes in $C_{T}$ using algorithm 1 and return and return 
		%\State Normalize $\bm{\hat{H}}_{T}$
		\State \Return Score vectors $(\bm{S}^{i} )_{1 \leq i \leq K}$%$\bm{C}$ (new candidate nodes)
		\Statex
		
	\end{algorithmic}
	
	%\vspace{-0.4cm}%
\end{algorithm}
\subsection{Second method : Graph similarity with kernels}

%This method relies on using another representation of entities after filtering.
The main idea of this method is to use another graph representation of entities, and use another similarity measure than scoring function $s$ used in equation $\ref{eq:graphFeatures}$ (e.g \textit{tfidf}).

\textbf{\textit{Graph of Words}:}
Graph of words (GoW) is a representation built over a sequence of objects in order to capture sequential relationships. It has proven its efficiency for several information retrieval problems \cite{Rousseau:2013:GTN:2505515.2505671}. Given a window size, nodes are added to the graph by their string representation. Edges are added between nodes in the same slidding window.  Figure \ref{GOW-example} shows an example with text (window size $=4$).

%	\begin{figure}[!h]
%			\begin{tabular}{cl}  
%			\begin{tabular}{c}
%				\includegraphics[height=4cm, width=3.5cm]{gow-example}
%			\end{tabular}
%			& \begin{tabular}{l}
%				\parbox{0.5\linewidth}{%  change the parbox width as appropiate
%					"Information   retrieval   is   the   activity   of   obtaining information  resources  relevant  to  an  information  need from a collection of information resources"
%				}
%			\end{tabular}  \\
%			
%		\end{tabular}
%		\caption{\label{GOW-example} Graph of word example on text data.}
%	\end{figure}

\begin{figure}[!h]
	\begin{minipage}{0.4\linewidth}
		\begin{tikzpicture}[every node/.style={sibling distance=30mm}]
		\begin{scope}[every node/.style={minimum size=0.1cm}]
		\def\words{{"obtaining","activity", "is","the"}}
		\foreach \i in {1, 2}
		\foreach \j in {4,5}
		\node [draw, fill=blue, circle, label=left:\scriptsize \pgfmathparse{\words[(\i-1)*2 + \j - 4]}\pgfmathresult] (n-\i\j) at (\i + 0.2*rand,\j + 0.2*rand)(\i\j){};
		\draw [->] (15) to [out=250,in=90] (14);
		\draw [->] (24) to [out=90,in=250] (25);
		\draw [->] (24) to [out=90,in=0] (15);
		\draw [->] (25) to [out=90,in=90] (15);
		
		\def\words{{"ressources","of", "retrieval"}}
		\foreach \i in {1,2,3}
		\node [draw, fill=blue, circle, label=above:\scriptsize \pgfmathparse{\words[(\i-1]}\pgfmathresult] (n-\i3) at (\i + 0.2*rand,3+ 0.2*rand)(\i3) {};
		\draw [->] (23) to [out=180,in=350] (13);

		\def\words{{"an","need", "from", "a"}}
		\foreach \i in {1, ..., 4}		
		\node [draw, fill=blue, circle, label=above:\scriptsize \pgfmathparse{\words[(\i-1]}\pgfmathresult] (n-\i1) at (\i + 0.2*rand,1 + 0.2*rand)(\i1){};
		\draw [->] (11) to [out=0,in=180] (21);
		\draw [->] (21) to [out=0,in=180] (31);
		\draw [->] (31) to [out=0,in=190] (41);
		\draw [->] (21) to [out=330,in=230] (41);
		
		\def\words{{"relevant","to", "information", "collection"}}
		\foreach \i in {1, ..., 4}
		\node [draw, fill=blue, circle, label=above:\scriptsize \pgfmathparse{\words[(\i-1]}\pgfmathresult] (n-\i2) at (\i + 0.2*rand,2 + 0.2*rand)(\i2) {};
		\draw [->] (12) to [out=0,in=180] (22);
		\draw [->] (22) to [out=0,in=180] (32);
		\draw [->] (42) to [out=180,in=0] (32);
		\draw [->] (32) to [out=120,in=60] (12);
		
		% Information
		\draw [->] (32) to [out=90,in=300] (13);
		\draw [->] (32) to [out=90,in=250] (33);
		\draw [->] (23) to [out=300,in=90] (32);
		\draw [->] (14) to [out=300,in=110] (32);
		\draw [->] (11) to [out=90,in=200] (32);
		\draw [->] (32) to [out=300,in=90] (21);
		\draw [->] (32) to [out=300,in=90] (31);
		
		% Activity
		\draw [->] (15) to [out=300,in=120] (23);
		
		% Relevant
		\draw [->] (12) to [out=200,in=130] (11);
		\draw [->] (13) to [out=250,in=90] (12);
		
		% Retrieval
		\draw [->] (33) to [out=40,in=300] (24);
		\draw [->] (33) to [out=30,in=300] (25);
		
		% Collection
		\draw [->] (41) to [out=30,in=300] (42);
		\draw [->] (42) to [out=10,in=70] (23);
		\draw [->] (31) to [out=10,in=210] (42);
		% An
		\draw [->] (22) to [out=210,in=60] (11);
		% Obtaining
		\draw [->] (14) to [out=250,in=90] (13);
		\draw [->] (23) to [out=130,in=270] (14);
		%the
		\draw [->] (25) to [out=0,in=0] (23);
		% Is
		%\foreach \i in {3,4,5}
		%	\node [draw, fill=blue, circle] (n-3\j) at (\i + 0.1*rand, 3 + 0.1*rand) {};
		%\foreach \i in {4,5}
		%	\foreach \j in {1,2,3,4}
		%		\node [draw, fill=blue, circle] (n-\i\j) at (\i + 0.1*rand,\j + 0.1*rand) {};
		%\foreach \i in {0,...,4}
		%\foreach \j [count=\jj] in {0,...,3}
		%\draw (n-\i\j) -- (n-\i\jj) (n-\j\i) -- (n-\jj\i);	
		\end{scope}
		\end{tikzpicture}
	\end{minipage}\hspace{1.5cm}\begin{minipage}{0.18\textwidth}"Information   retrieval   is   the   activity   of   obtaining information  resources  relevant  to  an  information  need from a collection of information resources"
	\end{minipage}
	\caption{\label{GOW-example} Graph of word example on text data.}
\end{figure}

For example, it has been shown \cite{Rousseau:2013:GTN:2505515.2505671} that k-core on the graph-of-words representation yield excellent keywords extraction.
Using this algorithm, is possible to compute a graph of word representation of a query and the definition of an entity.  

\subsubsection{ Graph similarity :} 

To compare these graph representations, several methods are available. We remind briefly two of them and compare query graph and select the most adapted to the \textit{NEL} problem.

\textbf{\textit{Sub-graph isomorphism}}	Given two graphs G and H input, and one must determining whether G contains a subgraph that is isomorphic to H is a way to determine proximity is a method to compare graph similarities \cite{cordella2004sub}.

\textbf{\textit{Graph kernels :} } Kernels have been popularized in the machine learning community as a powerful feature mapping tool, especially when combined with SVM classifiers. %A kernel has the fundamental property that it can be represented as an inner product on a Hilbert space. 
With graph structures, it is possible to define kernels that share same properties \cite{vishwanathan2010graph}. For implementation of several kernels, we refer to \cite{siglidis2018grakel}.

\textit{Example 1 : Shortest-path kernel	}

$ D(G) $ denotes the set of shortest distances between all node pairs in a graph G. The shortest-path kernel value on two given graphs $G_1 $ and $G_2$, is computed :
\begin{equation}
k(G_1,G_2) = \sum_{sd_i \in D(G_1)} \sum_{sd_j \in D(G_2)} \mathbbm{1}_{\{sd_i = sd_j\}}
\end{equation}

\textit{Example 2 :Pyramid match kernel:}	Pyramid match graph kernel uses a bag-of-vector representations of two given graphs. 
The idea of the algorithm is to map these vectors to multi-resolution histograms, and to compare these
histograms with a weighted histogram intersection measure
in order to find an approximate correspondence. For more details on this kernel, we refer to \cite{nikolentzos2017matching}.

To compare query and neighborhood graphs of words, we selected graph kernels for two reasons. First, graph kernels are offer lots of options due to various kernel definitions. Second, we conjectured (based on datasets adapted for named entity linking) that subgraph isomorphism is condition too strong for named entity linking. 

Given these definitions, our second method is obtained by adapting previous routine (eq. \ref{eq:graphFeatures}, and \ref{eq:graphLabels}, algorithm \ref{Algorithm:graph-mining-scoring}) replacing the scoring function $s$ by graph similarities into algorithm \ref{Algorithm:graph-mining-kernels}.  The final routine (inference) for both methods is summed up in algorithm \ref{Algorithm:KER for NEL}. 

\begin{algorithm}[!h]
	\small
	\caption{Graph of Words/Graph Kernels (GoW/GK) method}
	\label{Algorithm:graph-mining-kernels}
	
	\begin{algorithmic}[1]
		\Require{Same inputs as algorithm \ref{Algorithm:graph-mining-scoring}}
		%\State $\bm{C}_{0} \leftarrow \emptyset$
		%\State $\hat{\bm{H}}_{0} \leftarrow \emptyset $
		\For{$ i = 1$ to $K$}
		\parState{Get neighbor/close nodes in the graph of $e_i$}
		\parState{For each neighbor $n_j$, compute graph of word representation}
		\parState{Aggregate original entity filtering score and graph similarity scores between neighbors and the query graphs of words}
		\EndFor
		%\State Rank nodes in $C_{T}$ using algorithm 1 and return and return 
		%\State Normalize $\bm{\hat{H}}_{T}$
		\State \Return Score vectors $(\bm{S}^{i} )_{1 \leq i \leq K}$%$\bm{C}$ (new candidate nodes)
		\Statex
		
	\end{algorithmic}
\end{algorithm}

\vspace{-0.4cm}

\begin{algorithm}[!h]
	\small
	\caption{SGM and GoW/GK named entity identification (Inference)}
	\label{Algorithm:KER for NEL}
	\begin{algorithmic}[1]
		
		\Require{Knowledge base $\bm{B}$ and its graph $\bm{G}_{\bm{B}}$, mentions $(\bm{M}_i)_{1 \leq i \leq M}$, scoring threshold $\bm{K}$, trained predictor $\hat{\bm{F}}$ }
		\For{$ i = 1$ to $M$}
		\parState{Use algorithm 1 on mention $\bm{M}_i$ and $\bm{B}$, return a list of $\bm{K}$ top ranked entities $(\bm{E}^{1}_{h})_{1 \leq h \leq K }$ }
		\parState{Use algorithm 2 or 3 using $\bm{G}_{\bm{B}}$, on $\bm{K}$ entity candidates, return new score vectors}	
		%parState{Rank with $\bm{\alpha}$-convex combination of $\hat{\bm{H}}_{\epsilon}$ and scores}
		%\State Update $\bm{R}_{i}$ with concatenation of $\bm{E^{1}}$ and $\bm{E}^{2}$
		\parState{Evaluate $\hat{\bm{F}}$ on each vector score and use maximum a posteriori to infer estimated gold entity $\hat{\bm{G}}_{i}$}
		\EndFor
		\State \Return ${(\hat{\bm{G}}_{i})}_{1\leq i \leq M}$ (list of estimated gold entities)
		\Statex
		
	\end{algorithmic}
	
\end{algorithm}

~\\
\subsection{Computational complexity}
Our filtering algorithm time complexity is upper-bounded by $|\bm{M}||\bm{E}| = ME$. 
%(This bound can be improved considering the number of entity types). 
SGM procedure (algorithm \ref{Algorithm:graph-mining-scoring}) time complexity is proportional to $\mathcal{O}(M|T|K)$, where $|T|$ is the number of entity types considered, and $\bm{K}$ the remaining entity candidates.
Similarly, $G$ representing the kernel computational complexity, we can compute the complexity of our second method. In practice, $G$ is proportionnal to the number of words in the query times the number of words in an entity description, which is reasonable since entity descriptions and queries can be considered as short texts (less than thousand of words).
Therefore, the  \textit{first method (F+SGM)} complexity is: $\mathcal{O}(M(E+\bm{K}|T|))$, and the \textit{second Method (F+GoW/GK)} complexity is : $\mathcal{O}(M(E+\bm{K}|T|G))$. We report this in figure \ref{scalableIdentification}, along with some experimental computing times.

\section{Experimental setup and evaluation} \label{SEC:EXPERIMENTS}
%\subsection{Type features}
%\label{ssec:entityTypes}

%To identify entities type, we used a similar algorithm to \ref{Algorithm:scoring}.
%\subsection{Datasets : description}
\subsection{Configuration}

\begin{figure}[!h]
	\centering
	\renewcommand{\arraystretch}{1.2}
	\begin{tabular}{|l|c|c|c|c|}
		\hline
		\multirow{2}{2cm}{\textbf{Dataset}} & \multicolumn{3}{c|}{\textbf{Non-NIL mentions}} & \bf{Total}\\
		% \hline
		% \textbf{Inactive Modes} & \textbf{Description}\\
		\cline{2-5}
		& \textbf{PER} & \textbf{ORG} & \textbf{GPE} & ~ \\
		%\hhline{~--}
		\hline
		TAC09 (Test) & 255 & 1013 & 407  & 1675 \\ \hline
		TAC10 (Test) & 213 & 304 & 503 & 1020  \\ \hline
		CONLL (Test) & - & - & - & 4379  \\ \hline
		%\textbf{Total (Test)} & 468 & 1317 & 910 & 2695  \\ \hline
		\hline
		TAC10 (Train)& 335 & 335 & 404 & 1074 \\ \hline
		TAC14 (Train) & 1461 & 767 & 1313 & 3541 \\ \hline
		CONLL (Train + Valid) & - & - & - & 22516  \\ \hline
		%TRAIN 2015 & . & . & . & . \\ \hline
		%\textbf{Total train} & 1796 & 1102 & 1717 & 4615  \\ \hline		
		%Fourth condition & TBD & TBD & TBD & $\mu$ s  \\ \hline
	\end{tabular}
	\caption{\label{DatasetsDescription}Number of non-nil mentions in NIST TAC-KBP and CONLL Datasets}
\end{figure}
\textit{\textbf{Datasets :}}
%\textbf{to be completed}
%We split datasets in 3 entity types : Organization (ORG), People (PER), and Localization (GPE). 
Our datasets generated from COnLL and NIST TAC-KBP 2009-2010, contain for each query its gold entity id and type. %This gives 3 subsets on which we measure performance.

%We report here the results on the dataset provided by National institute of standards and technologies (NIST) for the named entity discovery challenge (TAC-KBP).
Figure \ref{DatasetsDescription} gives the number of samples for each dataset; more details are available on their respective official websites \footnote{\label{NIST-TAC}http://www.nist.gov/tac/}, \footnote{\label{NIST-TAC}https://www.mpi-inf.mpg.de/departments/databases-and-information-systems/research/yago-naga/aida/}.
%Github repository\textsuperscript{\ref{githubNelNote}}.
%\subsection{Evaluation: Results and comparison}

\textit{\textbf{NIL-detection}} : A \textit{nil} mention is mention that has no entity identified in the knowledge base.  As mentioned in the first part, we did not include in our work the problem of \textit{NIL-detection} (detection of entities out of the knowledge base). Following \cite{ganea2016probabilistic,ganea2017deep}, we removed NIL entities from the datasets. Therefore, performance comparison with other solutions (cf figure \ref{Results1}) has to take in account this feature.

\textit{\textbf{Implementation :}} Source code is available on demand.

\textit{\textbf{Entity types, ontology :}}
As discussed throughout the paper :
\begin{itemize}
	\item Our methods rely on a fined-grained classification of entities in the knowledge base. To generate fine-grained entity type inside the knowledge base, we crossed DBPedia with NIST TAC-KBP knowledge base using entity Wikipedia titles, and CoNLL with the 2016 Wikipedia Dump.
	\item We did not include fined-grained entity recognition on the queries : we suppose this given as input in the data.
\end{itemize}

\textit{\textbf{Graph kernels \& regressors:}} 
\begin{itemize}
	\item We report here results (figure \ref{Results1}, \ref{ImpactOfK}, \ref{scalableIdentification}) with the pyramid match graph kernel, for its low complexity among standard kernels \cite{nikolentzos2017matching}. We tried different graph kernels for our second method, including Shortest-path kernel, Weisfeiler-Lehman Kernel, and results were similar. 
	\item We tried several standard classifiers : regression trees, Support vector machines, and logistic regression. The results reported are obtained with logistic regression.
\end{itemize}

\textit{\textbf{Performance metrics:}}
A named entity linking system works as a search engine system, where we suppose there is only one relevant item (i.e the gold entity). Since we don't include NIL detection in our work, the most natural  performance measure of such a system is to evaluate the presence of the associated gold entity in the top K results. Averaging this quantity over the samples gives the precision or recall at Top K (same quantity since the gold entity is either in or out the top K). The main motivation to display results at K comes from industrial applications, where the top K entities would be human annotated to identify the good entity (examples : database matching, company identification). Therefore, in this context, Top-K precision is important. Finally, comparing a NEL scorer and NEL classifier using accuracy gives misleading results because of unbalanced classes.

\begin{figure*}
	%\centering
	\renewcommand{\arraystretch}{1.2}
	\begin{supertabular}{|p{0.15\linewidth}|p{0.1\linewidth}|p{0.1\linewidth}|p{0.1\linewidth}|p{0.1\linewidth}|p{0.1\linewidth}|p{0.1\linewidth}|p{0.1\linewidth}|}
		\hline 
		\multirow{2}{10cm}{\textbf{~}}   &\multirow{2}{10cm}{\textbf{Method }} & \multirow{2}{10cm}{\textbf{NIL\\detection}} & \multirow{2}{10cm}{\textbf{Fine-grained\\ entity types}} & \multirow{2}{10cm}{\textbf{Training \\samples nb.}}& \multicolumn{3}{c|}{\textbf{Average $\bm{P}@1$ (Accuracy) $\pm $ std in \%}} \\
		\cline{6-8}
		& & & & & \scriptsize{TAC09} & \scriptsize{TAC10} & \scriptsize{CONLL/AIDA}\\
		\hline
		Ganea, 2016 \cite{ganea2016probabilistic} & PGM & No & Not required & $\sim10^6, 10^7 $ &/ & / & $87.39$ \\
		Ganea, 2017 \cite{ganea2017deep} & PGM \& D.L & No & Not required & $\sim10^6, 10^7$ & / & / & $92.22$ \\
		Sun, 2015 \cite{sun2015modeling}  & D.L & No & Not required& $\sim10^6, 10^7$ & $82.26$ & $83.92$ & / \\
		Yamada, 2016 \cite{yamada2016joint}  &  D.L & No & Not required& $\sim10^6, 10^7$  & /  & $85.2$  & $93.1$ \\
		Yamada, 2017 \cite{yamada2017Learning} & D.L & No & Not required & $\sim10^6, 10^7$  & / & $87.7$ & $94.3$ \\ 
		Globerson, 2016 \cite{globerson2016collective}  & D.L & Yes & Not required&  $\sim10^6, 10^7$  &  / & $87.2$ & $92.7$\\
		Sil, 2017 \cite{sil2017neural}  & D.L & Not Detailed$^{\dagger}$ & Not required & $\sim10^6, 10^7$  &/  & $87.4$ & $93.0$ \\
		Raiman, 2018 \cite{raiman2018deeptype}   & D.L & Not Detailed$^{\dagger}$  & Required & $\sim10^6, 10^7$  & / & $90.85$ & $\bm{94.87}$\\
		Guo, 2011 \cite{guo2011graph}  & Graph-based & Yes & Not required & $\sim10^4$&  $84.89 $& $82.40$ & / \\
		Han, 2011 \cite{han2011collective}   & Graph-based & Yes & Not required & $\sim10^4$&  / & / & $81.91$ \\
		Hoffart, 2011 \cite{hoffart2011robust} & Graph-based & No & Not required& $\sim10^4$&  / & / & $81.91$ \\
		Usbeck, 2014 \cite{usbeck2014agdistis} & Graph-based & Yes & Not required& $\sim10^4$ & / & / & $73.0 $ \\
		\hline
		\hline
		%Filtering F &  $\bm{97.73}$  & $\bm{98.73}$ & $\bm{97.02}$ & $88.54$ & $87.06$ & $91.45 \pm 0.08$\\
		F+SGM & Graph-based &  No & Required & $\sim 10^3, 10^4$ & $ \bm{94.58 \pm 0.05}$ & $93.66 \pm 0.06$ &  $92.70 \pm 0.07$ \\
		F+GoW/GK & Graph-based & No &  Required & $\sim 10^3, 10^4$ & $ 93.67 \pm 0.06 $ & $\bm{94.70 \pm 0.05}$ & $93.56 \pm 0.06$\\\hline
	\end{supertabular}
	\caption{\label{Results1}Comparison of our apporach to state-of-the art methods with $\bm{K} = 7$.  For F+SGM, hyperparameter $\varphi$ is obtained with cross-validation. $^{\dagger}$Not detailed if NIL-detection method is not mentioned or explained. D.L stands for deep learning and PGM for probabilistic graphical model.}
	% \\(*) Authors claimed to obtain these values, but $\bm{K}$ is not given.}
	%. $A3$ parameters : $~ K=10$, $~ \epsilon = 10$, $\alpha=0.5$}
\end{figure*}

\textit{\textbf{Standard deviation, statistical significance:}}
We included standard deviation of the accuracy, but could not include p-significance of our method, due to the difficulty to reproduce other baselines experiments, namely :
\begin{itemize}
	\item Source code is not publicly available
	\item In case of deep learning methods, specific embeddings are not released
	\item Filtering method is not detailed
	\item Routine for tuning parameters is not explicit

\end{itemize}
%\hspace{-1cm}
\subsection{Results, comments and comparison}
\textit{\textbf{Results : }} We compare our methods with most performing baselines. Figure \ref{Results1} sums up our experimental results (averaged $\bm{P}@1$ is also referred as accuracy \cite{sun2015modeling}). 
%We report recall $\bm{R}@10$ and final precision $\bm{R}@1$ (also referred as accuracy \cite{sun2015modeling}). 
%Unfortunately, recall of, are not often completely reported, or not at all. In \cite{sun2015modeling}, somevalue is not presented, though the authors claimed to obtain $90.08\%$ and $91.17\%$ on TAC09 and TAC10 respectively  (cf \ref{Results1} and (*)).
Our method performs better than any existing graph-based methods. It outperforms all existing methods on two NIST TAC09 and TAC10, and is competitive with state-of-the arts methods on COnLL/AIDA.

\begin{figure}
	\centering
	%	\begin{minipage}
	\begin{tabular}{@{}c@{}}
		\begin{tikzpicture}[scale = 0.6]
		%(30,98.7) (50,99) 
		\begin{axis}[xmin=-1, xmax=22, xtick={1,5,10,15,20}, ymin=88, ymax=100, x dir=reverse, title={\textbf{COnLL/AIDA Dataset}},
		xlabel={K}, ylabel={Average accuracy (\%)}]
		%\addplot+[smooth,mark=*] plot coordinates
		%{ (1, 90)  (3, 92) (5,96.17) (10,97.6) (15, 97.83) (20,97.89)   };
		%\addlegendentry{Filtering}
		\addplot+[smooth,mark=x] plot coordinates
		{ (1, 90)(3, 91.70) (5,92.7) (10,92.7) (15, 92.2)  (20,92)   };
		\addlegendentry{Filtering + SGM}
		\addplot+[smooth,mark=x] plot coordinates
		{ (1, 90) (3, 92.56) (5,93.56) (10,93.56) (15, 92.8) (20,92) };
		\addlegendentry{Filtering + GK}
		\end{axis}
		\end{tikzpicture}
	\end{tabular}
	\begin{tabular}{@{}c@{}}
		\begin{tikzpicture}[scale=0.50]
		
		%(30,98.7) (50,99) 
		\begin{axis}[xmin=-1, xmax=22, xtick={1,5,10,15,20}, ymin=90, ymax=100, x dir=reverse, title={\textbf{NIST TAC-KBP 2009}},
		xlabel={K}, ylabel={Average accuracy (\%)}]
		%\addplot+[smooth,mark=*] plot coordinates
		%{ (1, 91)  (3, 94) (5,96.17) (10,97.6) (15, 97.83) (20,97.89)   };
		%\addlegendentry{Filtering}
		\addplot+[smooth,mark=x] plot coordinates
		{ (1, 91)(3, 91.5) (5,94.58) (10,94.58) (15, 94)  (20,93.2)   };
		\addlegendentry{Filtering + SGM}
		\addplot+[smooth,mark=x] plot coordinates
		{ (1, 91) (3, 92) (5,93.67) (10,93.67) (15, 93) (20,92.7) };
		\addlegendentry{Filtering + GK}
		\end{axis}
		\end{tikzpicture}
	\end{tabular}
	\begin{tabular}{@{}c@{}}
		\begin{tikzpicture}[scale=0.5]
		
		%(30,98.7) (50,99) 
		\begin{axis}[xmin=-1, xmax=22, xtick={1,5,10,15,20}, ymin=90, ymax=100, x dir=reverse, title={\textbf{NIST TAC-KBP 2010}},
		xlabel={K}]
		%ylabel={Average accuracy (\%)}
		%\addplot+[smooth,mark=*] plot coordinates
		%{ (1, 92.35)  (3, 93) (5,96.17) (10,97.6) (15, 97.83) (20,97.89)   };
		%\addlegendentry{Filtering}
		\addplot+[smooth,mark=x] plot coordinates
		{ (1, 92.35) (3, 92.5) (5,93.66) (10,93.5) (15, 93.1)  (20,93)   };
		\addlegendentry{Filtering + SGM}
		\addplot+[smooth,mark=x] plot coordinates
		{ (1, 92.35) (3, 92.7) (5,94.7) (10,94.5) (15, 94.2) (20,94) };
		\addlegendentry{Filtering + GK}
		\end{axis}
		\end{tikzpicture}
	\end{tabular}
	%		\end{minipage}
	\caption{\label{ImpactOfK} Impact of K on average $\bm{P}@1$.}
\end{figure}
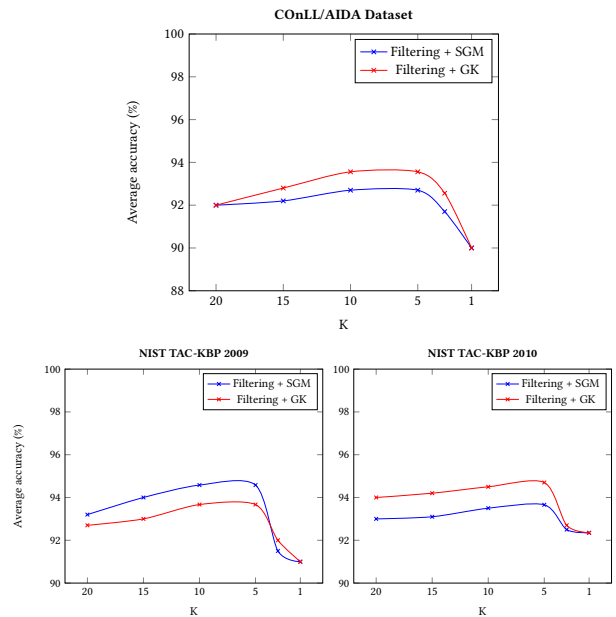

%hese datasets.
% It improves in a significant amount on CONLL/AIDA dataset.%Results \ref{NISTResults1} show that graph mining, after scoring, improves accuracy while name and context scoring fail.
%\textit{\textbf{Interpretation : }} : 
\textit{\textbf{Impact of parameter K : }} We report impact of parameter K on final average $\bm{P}@1$. Results are shown on figure \ref{ImpactOfK}. The curves show an experimental trade-off between exploration and a strict candidate filtering. Low values of $\bm{K}$ don't allow enough entity exploration and cause a drop in precision. On the contrary, high values of K yields too many entity candidates. Results are similar for $ 5 \leq K  \leq  10$. 
%\tikzsetnextfilename{Figs/pgfBasicPlot}

\textit{\textbf{Scalability :}}
Figure \ref{scalableIdentification} sums up asymptotic computational complexities and experimental computing time on a sub-instance of COnLL/AIDA (1000 queries, 2.8 million entities). These times are only indicative, since there is room for improvement, especially due to the choice of the language (here, Python) and code optimization. 
\begin{figure}
	\centering
	\renewcommand{\arraystretch}{1.2}
	\begin{tabular}{|l|c|c|c|c|c|}
		\hline
		\multirow{3}{1cm}{\textbf{Method}} & \multirow{3}{2.2cm}{\textbf{Computational \\ complexity}}  & \multicolumn{4}{c|}{\textbf{Experimental time (mn.)}}\\
		\cline{3-6}
		& & \multicolumn{2}{c|}{\textbf{Setup 1}} & \multicolumn{2}{c|}{\textbf{Setup 2}}\\
		\cline{3-6}
		&  &  $\bm{K} = 5$& $\bm{K} = 20$ & $\bm{K} = 5$& $\bm{K} = 20$  \\
		\hline
		F  & $\mathcal{O}(ME)$ & 620 & 620  & 44 & 44 \\  \hline
		SGM & $\mathcal{O}(M\bm{K}|T|)$ & 240 & 290 & 14 & 16 \\ \hline
		GoW/GK & $\mathcal{O}(M\bm{K}|T|G)$ & 2642 & 306   & 16 & 20 \\ \hline
	\end{tabular}
	\caption{\label{scalableIdentification} \small{Scalable entity identification. Computing times rounded to the minute. $ M = 1000  \text{, }E = 2.8\times 10^6$, $G \leq 200$, $T = 250$. 
			Setup 1 : Single CPU with 32Gb Ram, Intel(R) Xeon(R) CPU E5-2407 4-cores 2.40GHz.
			Setup 2 : Distributed cluster with variety of 20 CPU processors equivalent to setup 1. (using Spark/Hadoop technology)}}
\end{figure}

\textit{\textbf{Comparison : advantages and limitations : }} 
Our methods yields remarkable precision $\bm{P}@1$ on TAC09 dataset, CONLL/AIDA and TAC10 datasets. 
For features extraction, the number of parameter that need to be tuned is reasonable, namely :
\begin{itemize}
	\item F+SGM :  the type mapping function $\varphi$ (obtained with cross validation) and $\bm{K}$
	\item  F+GoW/GoK : graph of words window size, and $\bm{K}$.~\\
\end{itemize}
%or the first method, the type mapping function $\varphi$ obtained with cross validation. 

For named entity discovery, our methodology has two limitations. First, we did not include NIL-detection. Second, our filtering method depends on fine-grained classification of named entities (example : GPE: Cities, Countries, ORG: Company, SoccerClub, PER : Actor, Singer, Politician, ...). We supposed such classification was available in the query, whereas progress stil has to be made to tag named entities with fine-grained classification \cite{ling2012fine}.

%\newpage
\section{Conclusion} 
In this paper, we proposed a new methodology concerning the problem of named entity linking. Capitalizing on experimental factors of entity mis-identification, we first proposed a filtering algorithm based on standard information retrieval techniques. Then, each entity candidate is matched with new features built on a knowledge subgraph centered on their corresponding node. Our methods perform individual linking  : mentions are considered separately. Eventhough we did not include NIL detection or fined-grained entity recognition, we have shown empirically that our graph-based named entity identification outperforms state-of-the-art methods on two datasets and is competitive on one dataset. 

We have also show that our filtering and graph-mining features extraction scales well : their computational complexity is linear with respect to the numbers of queries and entities, and they have good experimental computing time for short text documents.

There are some advantages of our method over deep learning approaches. First, entity features are interpretable. Second, our linking system is relatively easy to implement in a real system, with relatively few hyperparameters, especially for the second method using graph kernels. Last but not least, it does not require lots of data to reach good experimental performance. Indeed, only a few thousands of training samples were used to reach these results.

We hope this work will serve as a baseline for named entity linking when fine-grained entity ontology is available. This work also invites us to complete it with graph based named entity type classification. Moreover, we could potentially improve performance with careful attention given to a new graph kernel for named entity linking. We leave these ideas for future work.
% $\mathcal{O}(n)$ with logistic regression
%\section*{Acknowledgements}
%We would like to thank all Dascim LIX team members for their help, especially Christos Giatsidis, and Christos Xypolopoulos concerning the use of Hadoop and Spark technologies. Also, thanks go to Octavian Ganea (ETH Zurich) for his time and constructive discussions. Finally, thanks go to Geoffrey Scoutheeten (BNP Paribas), for his help on the final reading. This work was partially supported by by Association Nationale de la recherche et de la technologie (ANRT) and BNP Paribas under the Cifre convention 2016/2017. %\cite{ganea2016probabilistic}.

\bibliographystyle{ACM-Reference-Format}
\bibliography{scalable-igraph-based-nei-WWW19}

\end{document}